\documentclass[twocolumn]{aastex631}
\usepackage{natbib}
\usepackage{amsmath}
\usepackage[T1]{fontenc}
\usepackage[utf8]{inputenc}

\def\h{\hskip -0.3 mm}
\def\nd{\nodata}

\def\apjl{{ApJL}}
\def\apjs{{ApJS}}

\def\e#1{$\times$10$^{#1}$}

\def\farcs{\hbox{$.\mkern-4mu^{\prime\prime}$}}

\def\hal{H$\alpha$}
\def\hb{H$\beta$}

\def\lax{{$\mathrel{\hbox{\rlap{\hbox{\lower4pt\hbox{$\sim$}}}\hbox{$<$}}}$}}
\def\gax{{$\mathrel{\hbox{\rlap{\hbox{\lower4pt\hbox{$\sim$}}}\hbox{$>$}}}$}}
\def\simlt{\lower.5ex\hbox{$\; \buildrel < \over \sim \;$}}
\def\simgt{\lower.5ex\hbox{$\; \buildrel > \over \sim \;$}}

\def\mbh{{$M_{\rm BH}$}}

\def\cm2{cm$^{-2}$}

\def\oiii{[\ion{O}{3}]}

\def\lbol{$L_{{\rm bol}}$}

\def\l5100{$L_{5100}$}
\def\ll5100{$\log L_{\rm 5100}$}
\def\lloiii{$\log L_{\rm [O\ III]}$}
\def\llmbh{$\log M_{\rm BH}$}
\def\-->{$\rightarrow$}

\shorttitle{MIR Structure Function of AGNs}
\shortauthors{Son et al.}

\begin{document}

\title{The Structure Function of Mid-infrared Variability in Low-redshift Active Galactic Nuclei}

\author[0000-0002-5346-0567]{Suyeon Son}
\affiliation{Department of Astronomy and Atmospheric Sciences,
Kyungpook National University, Daegu 702-701, Korea}

\author[0000-0002-3560-0781]{Minjin Kim}

\affiliation{Department of Astronomy and Atmospheric Sciences, 
Kyungpook National University, Daegu 702-701, Korea}

\author[0000-0001-6947-5846]{Luis C. Ho}

\affiliation{Kavli Institute for Astronomy and Astrophysics, Peking 
University, Beijing 100871, China}

\affiliation{Department of Astronomy, School of Physics, Peking University, Beijing 100871, China}

\correspondingauthor{Minjin Kim}
\email{mkim.astro@gmail.com}

\begin{abstract}
Using the multi-epoch mid-infrared (MIR) photometry from the Wide-field Infrared Survey Explorer spanning a baseline of $\sim10$ yr, we extensively investigate the MIR variability of nearby active galactic nuclei (AGNs) at $0.15 < z < 0.4$. We find that the ensemble structure function in the W1 band ($3.4\ \mu$m) can be modeled with a broken power law. Type 1 AGNs tend to exhibit larger variability amplitudes than type 2 AGNs, possibly due to the extinction by the torus. The variability amplitude is inversely correlated with the AGN luminosity, consistent with a similar relation known in the optical. Meanwhile, the slope of the power law increases with AGN luminosity. This trend can be attributed to the fact that the inner radius of the torus is proportional to the AGN luminosity, as expected from the size$-$luminosity relation of the torus. Interestingly, low-luminosity type 2 AGNs, unlike low-luminosity type 1 AGNs, tend to exhibit smaller variability amplitude than do high-luminosity AGNs. We argue that either low-luminosity type 2 AGNs have distinctive central structures due to their low luminosity or their MIR brightness is contaminated by emission from the cold dust in the host galaxy. Our findings suggest that the AGN unification scheme may need to be revised. We find that the variability amplitude of dust-deficient AGNs is systematically larger than that of normal AGNs, supporting the notion that the hot and warm dust in dust-deficient AGNs may be destroyed and reformed according to the strength of the ultraviolet radiation from the accretion disk.   
\end{abstract}


\section{Introduction} 

Active galactic nuclei (AGNs) radiate strong multi-wavelength continuum emission originating from complex structures around supermassive black holes (BHs), such as the accretion disk (AD), corona, jet, and dusty torus. One of the unique features of AGN emission is its variability over a wide range of time scales, indicating that the emission arises over a large range of physical scales \cite[e.g.,][]{matthews_1963}. Therefore, the variability can be used to probe the physical properties of central structures in AGNs, which, apart from interferometric observations  \cite[e.g.,][]{swain_2003, gravity_2020a, gravity_2020b}, are barely resolved with conventional imaging data. Variability has also been widely used for AGN selection \cite[e.g.,][]{vandenbergh_1973, choi_2014, burke_2023}. While the physical origin of the variability remains under debate \cite[e.g.,][]{lyubarskii_1997, ulrich_1997, kawaguchi_1998, livio_2003, dexter_2011, kubota_2018, sun_2020}, the X-ray and UV/optical variability can be described by a stochastic process \cite[e.g.,][]{peterson_1997, kelly_2011}. More specifically, the UV/optical continuum is well modeled by a damped random walk, for which the shape of the power spectral density is a broken power law (e.g., \citealt{kelly_2009, macleod_2010, tang_2023}; but see \citealt{kasliwal_2015}).

The power spectral density of the light curves in the UV/optical band can be characterized by the variability amplitude, a power-law index of approximately $-2.0$ at high frequencies, and a characteristic frequency below which the power spectral density flattens. Various studies argued that these parameters may be related to the physical properties of AGNs (e.g., wavelength, BH mass, Eddington ratio, and AGN luminosity; \citealt{vandenberk_2004, li_2018, suberlak_2021, tang_2023}). However, different studies reached different conclusions regarding the dependence of the variability properties on the AGN properties, possibly attributed to the bias present in constructing the power spectral density introduced by sample selection, insufficient cadence, length of the time-series data, and method used to construct the power spectral density \citep{kozlowski_2017,suberlak_2021}. For example, \citet{li_2018} demonstrated, using ground-based survey data and ensemble analysis, that the variability amplitude is inversely correlated with the AGN luminosity but is independent of the BH mass (see also \citealt{kelly_2009}). However, \citet{macleod_2010} argued that the variability amplitude is correlated with the BH mass and inversely correlated with the Eddington ratio. In addition, the power-law slope at high frequencies is reported to be a constant (approximately $-2$) using datasets obtained from ground-based telescopes that may suffer from a sparse sampling rate. On the contrary, studies with high-cadence light curves obtained from the Kepler mission found that the power-law slope is not a constant and is correlated with the AGN properties \cite[e.g.,][]{mushotzky_2011, smith_2018}.       

While the variability of the continuum from the innermost regions of AGNs has been extensively studied with multi-epoch X-ray, UV/optical, and radio observations, the variability in the mid-infrared (MIR) has yet to be investigated with a large dataset \cite[e.g.,][]{kozlowski_2016a, son_2022, li_2023}. One of the obstacles toward realizing this goal is the lack of MIR time series owing to the relative difficulty of obtaining such data from ground-based telescopes. As the MIR continuum comprises AD emission reprocessed by the dust in the torus, it carries information not only on the intrinsic characteristics of the light from the AD but also helps us to constrain the structural properties of the torus \cite[e.g.,][]{kawaguchi_2011, son_2022, li_2023}. In this regard, it is instructive to indirectly probe the continuum from the AD of type 2 sources to test the unification model of AGNs. Despite the importance of MIR variability in understanding the structure of the dusty torus, there have been few quantitative examinations of this variability \cite[e.g.,][]{Kozlowski_2010, kozlowski_2016a, sanchez_2017, wang_2020}.

\begin{figure*}[t!]
\centering
\includegraphics[width=0.98\textwidth]{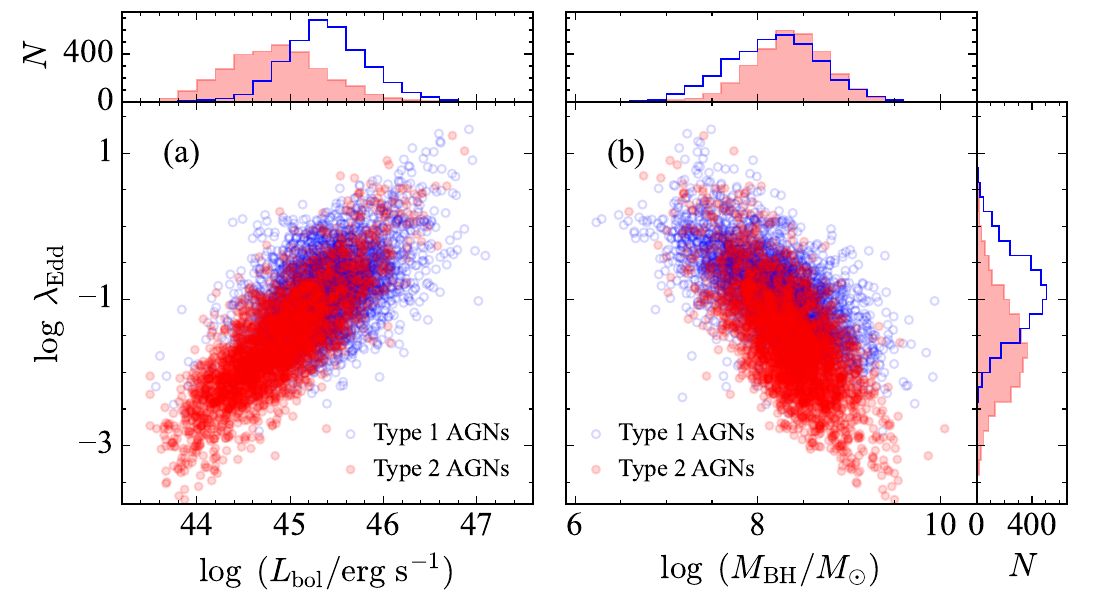}
\caption{
Distribution of Eddington ratio versus (a) bolometric luminosity and (b) BH mass for our sample of AGNs. The blue circles and open histogram denote type 1 AGNs, while filled red circles and filled histogram represent type 2 AGNs. 
}
\end{figure*}

Owing to the continuous survey conducted by the Wide-field Infrared Survey Explorer (WISE), $\sim 10$ years of all-sky data are available, which permits systematic variability studies of AGNs to be conducted, albeit with a sparse cadence ($\sim6$ months; \citealt{wright_2010}). The WISE database, in combination with complementary optical data, can be used to advance a variety of studies on AGN variability, including reverberation mapping, searching for changing-look AGNs or tidal disruption events, and investigating the physical properties of the hot dust components \cite[e.g.,][]{lyu_2019,sheng_2020, yang_2020,jiang_2021,son_2022b, li_2023}. Indeed, these data are also ideal for studying the properties of the MIR variability for a large sample of AGNs.   

To explore the characteristics of the MIR variability of nearby AGNs, we investigate the variability structure function (SF) of multi-epoch MIR data for a large sample of nearby AGNs obtained with WISE spanning a baseline of $\sim10$ yr, paying careful attention to treatment of noise estimation and host galaxy subtraction. In Section 2, we describe our sample selection and the method used to construct the MIR light curves. Section 3 introduces detailed methods for computing robust ensemble SFs. The SFs of various subsamples are presented in Section 4. The physical origins of the SF variations are discussed in Section 5. Finally, the results are summarized in Section 6. Throughout the paper, all magnitudes refer to the Vega system, and we adopt the cosmological parameters $H_0=100h=67.4$ km ${\rm s}^{-1}$ ${\rm Mpc}^{-1}$, $\Omega_m=0.315$, and $\Omega_\Lambda=0.685$ (\citealt{planck_2020}).

\section{Sample and Data}

For type 1 AGNs, the parent sample is drawn from the Data Release 14 quasar catalog of the Sloan Digital Sky Survey (SDSS; \citealp{paris_2018}). Because the light contribution from the host galaxy is significant in the W1 band ($3.4$ $\mu$m), careful subtraction of the host component is necessary \cite[][]{son_2022,son_2023}. We constrain the host magnitude by fitting the spectral energy distribution (SED). We impose a low-redshift cut ($z > 0.15$) to mitigate significantly extended sources, whose photometric data among the different datasets can be inconsistent due to mismatches in the photometric method and spatial resolution. At $z \approx  0.15$, the point-spread function (PSF) full-width at half maximum (FWHM) of $6\arcsec$ in the W1 band corresponds to $\sim 15$ kpc, which ensures that most of the galaxy flux is included in the WISE photometry. To facilitate more effective estimation of the host galaxy contribution using rest-frame near-infrared data, we only choose type 1 AGNs with 2MASS \citep{skrutskie_2006} counterparts with $z<0.4$, which additionally ensures that the \hal\ region is covered by the SDSS spectra for AGN classification. These redshift cuts are also crucial for minimizing the effect of cosmic evolution on the dust properties of the torus and for neglecting possible dependences on rest-frame wavelength \cite[e.g.,][]{kawaguchi_2011}. A total of 4295 type 1 AGNs are initially chosen.           

We use the same redshift cut ($0.15 < z<0.4$) to select 6854 type 2 AGNs from the SDSS Data Release 8, based on spectral classification from the Baldwin, Philips \& Terlevich diagram (i.e., ``bptclass''=4; \citealp{baldwin_1981, veilleux_1987}) provided by the Max Planck Institute for Astrophysics and the Johns Hopkins University (MPA-JHU) value-added catalog\footnote{\url{https://www.sdss3.org/dr10/spectro/galaxy_mpajhu.php}} \cite[][]{brinchmann_2004}. We exclude low-ionization nuclear emission-line regions (LINERs; \citealp{heckman_1980}), whose extremely low luminosities and Eddington ratios \citep{ho_2008,ho_2009} preclude accurate photometry of the nucleus based on the relatively low-resolution WISE images. For type 2 AGNs, matching with 2MASS is unnecessary because the total stellar masses from the MPA-JHU catalog are based on optical photometric data \cite{kauffmann_2003}.

\begin{figure*}[htp]
\centering
\includegraphics[width=0.98\textwidth]{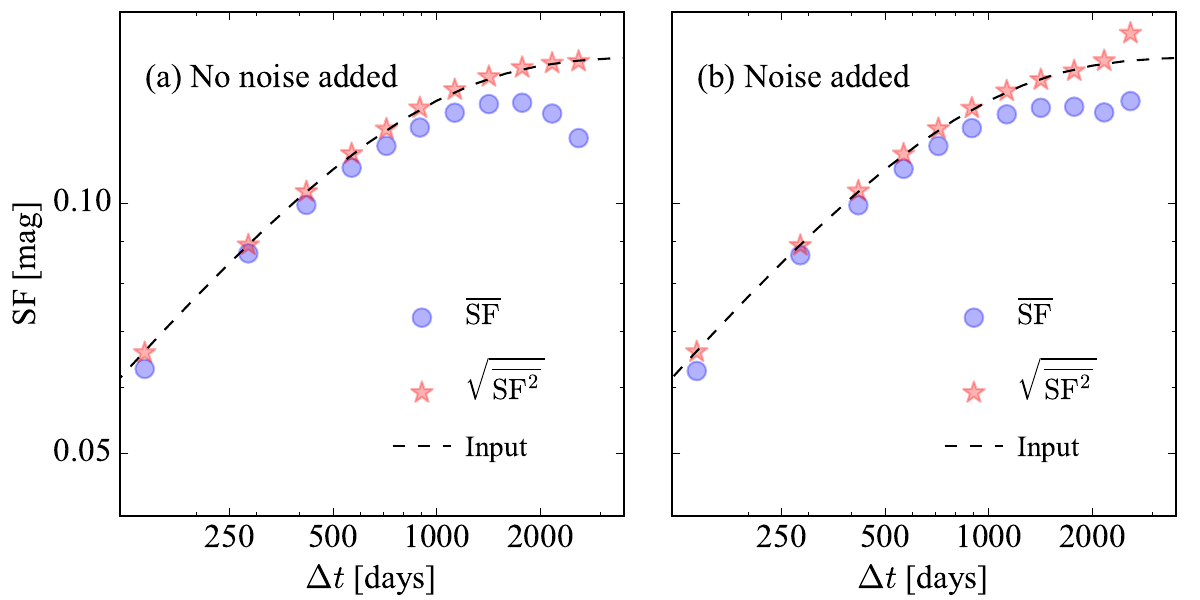}
\caption{
Ensemble SF estimated from light curves simulated (a) with no noise added and (b) with noise added. The dashed line is the input SF. The blue circles and red stars denote the mean of the SF and the square root of the mean SF, respectively. 
}
\end{figure*}

To investigate the SF of SDSS AGNs in the MIR, we use the multi-epoch data provided by the Near-Earth Object Wide-field Infrared Survey Explorer (NEOWISE), which spans a baseline of $\sim10$ yr from December 23, 2013 to December 13, 2022 \citep{mainzer_2011}. In view of the typical positional uncertainty of the WISE dataset ($\sim0\farcs5$; see also \citealp{assef_2013, son_2022})\footnote{\url{https://wise2.ipac.caltech.edu/docs/release/allwise/expsup}}, we adopt a radius of $2 \arcsec$ to cross-match between SDSS and NEOWISE. We only use time-series data from the W1 band because its signal-to-noise ratio is substantially higher than that from the W2 band ($4.6$ $\mu$m). To secure a robust analysis of the SF, we only consider samples with more than 17 epochs in the W1 band. Finally, to exclude non-variable objects, which are not appropriate for further analysis, we employ variability probability criterion of $P_{\rm var} > 0.95$ \citep{mclaughlin_1996, sanchez_2017}, where $P_{\rm var}=1-Q$ and $Q$, the probability that the observed light curve originates from a non-variable object, is estimated using the $\chi^2$ distribution at a given degree of freedom ($N-1$, with $N$ the number of observing epochs). A high $P_{\rm var}$ (low $Q$) indicates that an object is likely variable. The final sample contains 3506 type 1 and 3074 type 2 AGNs. 

As NEOWISE is designed to survey the entire sky over a six-month interval, any given field is observed $\sim13-14$ times on average over a few days during each visit. We retrieve the photometric measurements from the single exposures of each visit and compute a mean value with a $3 \sigma$ clipping. To discard suspicious measurements, we only use photometric data flagged with $cc\_flags = 0$, $qual\_frame > 0$, $qi\_fact > 0$, $saa\_sep > 0$, and $moon\_masked = 0$. The method from \citet{lyu_2019} is adopted initially to calculate the uncertainty of the W1 magnitude:

\begin{equation}
\sigma_{\rm in}^2 = \frac{1}{N_s-1}\sum_{i=1}^{N_s} {(m_{i}-
\overline{m_\mathrm{epoch}})^2} + \frac{1}{N_s^2}\sum_{i=1}^{N_s} 
\sigma_{i,\mathrm{pho}}^2 + \frac{1}{N_s}\sigma_\mathrm{sys}^2,    
\end{equation}

\noindent
where $N_s$ is the number of single exposures in each epoch, $m_i$ and $\sigma_{i,\mathrm{pho}}$ represent the magnitude and uncertainty of a single exposure, $\overline{m_\mathrm{epoch}}$ is the mean magnitude in each epoch, and $\sigma_\mathrm{sys}$ is the systematic uncertainty ($\sim0.016$ mag) due to the instability of the system. However, from extensive testing of the SF from non-variable sources, we find that the initial uncertainty can be substantially overestimated (see Section 3.3.2), and the final uncertainty is estimated better by dividing $\sigma_{\rm in}$ by $\sqrt{N_s}$ ($\sigma_e = \sigma_{\rm in} / \sqrt{N_s}$).

We extract various AGN properties (Figure~1) derived from the spectral measurements of \cite{rakshit_2020} to explore the physical connection between the SF and properties of the AGN. The BH mass (\mbh) for type 1 AGNs is derived using the virial method: \mbh $= f Rv^2/G$, where $R$ is the size of the broad-line region inferred from the continuum luminosity at 5100~\AA\ (\l5100) using the size$-$luminosity relation (\citealp{kaspi_2000, bentz_2013}, $v$ the velocity width of broad \hb\ emission, and $f$ is a scaling factor determined by the geometry and kinematics of the line-emitting region. We adopt the BH mass estimator from \citet[][]{ho_2015} appropriate for both bulge types (see \citealp{ho_2014} on the systematic difference of $f$ for classical and pseudo bulges): $M_{\rm BH} = 10^{6.91}({\rm FWHM}/1000\ {\rm km\ s^{-1}})^2(L_{5100}/10^{44}\ {\rm erg\ s^{-1}})^{0.533}\, M_\odot$. For type 2 AGNs, we utilize the empirical relation between BH mass and the total stellar mass of the host ($M_*$), as calibrated by \cite{greene_2020} for all (early and late) galaxy types: $M_{\rm BH} = 10^{7.43}(M_*/3\times10^{10}\,M_\odot)^{1.61}\, M_\odot$. The stellar masses, extracted from the MPA-JHU catalog, are derived by fitting the spectral energy distribution from the SDSS photometry with the stellar population model. The BH mass measurements have uncertainties of $\sim0.5$ dex and $\sim0.8$ dex for type 1 and type 2 AGNs, respectively. We use the \oiii\ luminosity to trace the strength of both AGNs types, and, as an additional check, \l5100\ for type 1 sources alone. Converting the \oiii\ luminosity to the bolometric luminosity requires consideration of possible correction for dust extinction \cite[see][for an extensive discussion of the bolometric conversion]{kong_2018}. For our sample of type 1 AGNs, \l5100\ correlates more strongly with the observed \oiii\ emission than with the line luminosity after extinction correction based on the Balmer decrement. We therefore choose the bolometric correction of \cite{heckman_2004}, $L_{\rm bol}=3500 L_{\rm [O\ III]}$, which is based on the observed \oiii\ luminosity. The Eddington ratio $\lambda_{\rm Edd}\equiv L_{\rm bol}/L_{\rm Edd}$, where the Eddington luminosity $L_{\rm Edd}=1.26 \times 10^{38} \,M_{\rm BH}/M_\odot$.

\begin{figure}[t!]
\centering
\includegraphics[width=0.45\textwidth]{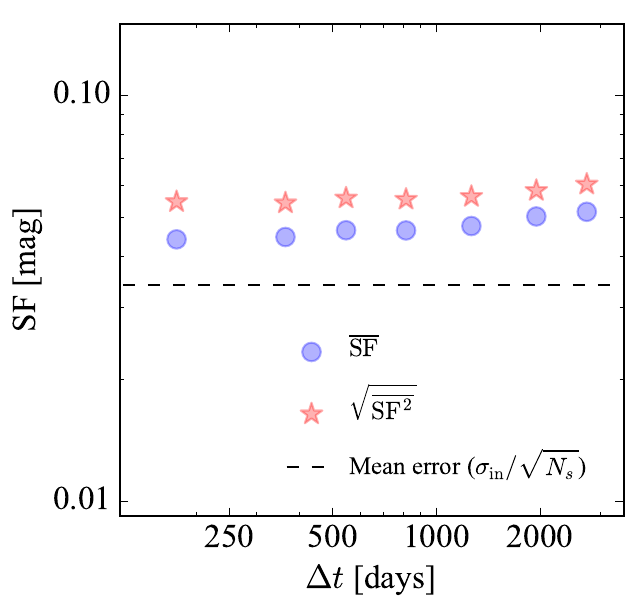}
\caption{
Ensemble SF of non-variable sources. The dashed line denotes the mean error of the sources. The blue circles and red stars denote the mean of the SF and the square root of the mean SF, respectively.
}
\end{figure}

\section{Structure Function}

\subsection{Definition}

AGN variability can be modeled with the damped random walk model, wherein the variability amplitude of the light curve as a function of a time lag obeys a power law on short time scales and flattens on longer time scales \cite[e.g.,][]{kelly_2009}. Alternatively, the variability can be modeled non-parametrically with the SF, which is defined as the mean magnitude difference (SF) as a function of time lag ($\Delta t$; e.g., \citealt{simonetti_1985,kawaguchi_1998,kozlowski_2016b}):
\begin{equation}
\begin{split}
    {\rm SF}^2(\Delta t) = \frac{1}{N_{\Delta t, {\rm pair}}} \sum_{i=1}^{N_{\Delta t, {\rm pair}}} (m(t) - m(t+\Delta t))^2 \\ - (\sigma_{\rm e}^2(t) + \sigma_{\rm e}^2(t+\Delta t)), 
\end{split}
\end{equation}
\noindent
where $N_{\Delta t, {\rm pair}}$ denotes the number of pairs associated with $\Delta t$, $m$ is the observed magnitude, and $\sigma_{\rm e}$ represents the uncertainty in each epoch. As expected from the damped random walk model, the SF in the optical band is divided into two power laws, ${\rm SF} \propto \Delta t^\gamma$ at a break time scale $t_{\rm break}$, such that: $\gamma \approx 1$ for $\Delta t \leq t_{\rm break}$ and $\gamma \approx 0$ for $\Delta t < t_{\rm break}$. 

Ideally, $\sigma_{\rm e}$ can be estimated from the standard deviation of the multi-epoch photometric data [i.e., ${\rm SF}(0)$] for the non-variable objects and is known to be independent of $\Delta t$. Alternatively, the dependence on $\Delta t$ can be computed from the SF derived from the light curves of non-variable sources as $ {\rm SF}^2_{\rm non-variable}(\Delta t) = \sigma_{\rm e}^2(t) + \sigma_{\rm e}^2(t+\Delta t)$ (\citealp{kozlowski_2016a}). As described in Section 3.3.2, we evaluate the ensemble SF using the WISE light curves from non-variable sources and find that the SF depends on $\Delta t$ as ${\rm SF}(\Delta t) = {\rm SF}(0) (0.999+0.055\times \Delta t/1000)$. We use this relation to estimate the SF. Note that the observed $\Delta t_{\rm obs}$ is converted to the rest frame, namely $\Delta t = \Delta t_{\rm obs} / (1+z)$. 

Owing to the sparse cadence of the NEOWISE data ($\sim 6$ months), it may not be ideal to use the SF from individual objects to explore the statistical properties of the MIR variability. Instead, we utilize an ensemble SF, which is computed by averaging the SFs for a given $\Delta t$ obtained from subsamples with similar AGN properties (e.g., AGN type, AGN luminosity, BH mass, and dust properties; \citealp{almani_2000, vandenberk_2004, li_2018}).

\subsection{Host Subtraction}

The contribution of the host galaxy can be significant in the W1 band and should be carefully removed to yield a robust estimate of the AGN luminosity. The SF, therefore, is highly sensitive to host subtraction. For example, over-subtraction of the host light can lead to an overestimation of the SF. For type 1 AGNs, we perform a fit to the SED spanning from the optical to the MIR, using integrated photometry derived from SDSS, 2MASS, and WISE, as detailed in \citet{son_2023}. We consider two components, one for the host and the other for the AGN. The AGN component is represented by three SEDs---hot dust-deficient (HDD), warm dust-deficient (WDD), and normal AGNs---empirically determined based on the presence of hot dust near $1-3$ $\mu$m and warm dust around $3-10$ $\mu$m. While the HDD AGNs are defined as lacking both hot and warm components in their SED, the SED of normal AGNs is well fit with the template SED of bright QSOs from \citet{elvis_1994}. The host component is modeled with seven templates: an old stellar population with 7 Gyr old stars and empirical templates of inactive galaxies (Hubble type E, S0, Sa, Sb, Sc, and Sd) from the Spitzer Wide-Area Infrared Extragalactic Survey library \citep{polletta_2007}. 

For type 2 AGNs, we find that SED fitting does not provide satisfactory results, possibly because the extended morphology of these sources may introduce systematic offsets in the photometry among the different datasets. \citet{liho_2023} demonstrate, for instance, that the photometry provided by the 2MASS and WISE catalogs can be systematically underestimated if the sources are extended. Therefore, we instead estimate the luminosity expected in the W1 band based on the stellar mass derived from the SDSS photometry, in combination with the mass-to-light ratio in the W1 band adopted from \citet{kettlety_2018}. The average flux fraction of the host in the W1 band is $0.20\pm0.12$ for type 1 AGNs and $0.47\pm0.24$ for type 2 AGNs. Because of their lower bolometric luminosity, type 2 AGNs have significantly more dominant hosts than type 1 AGNs.

\begin{figure}[t!]
\centering
\includegraphics[width=0.45\textwidth]{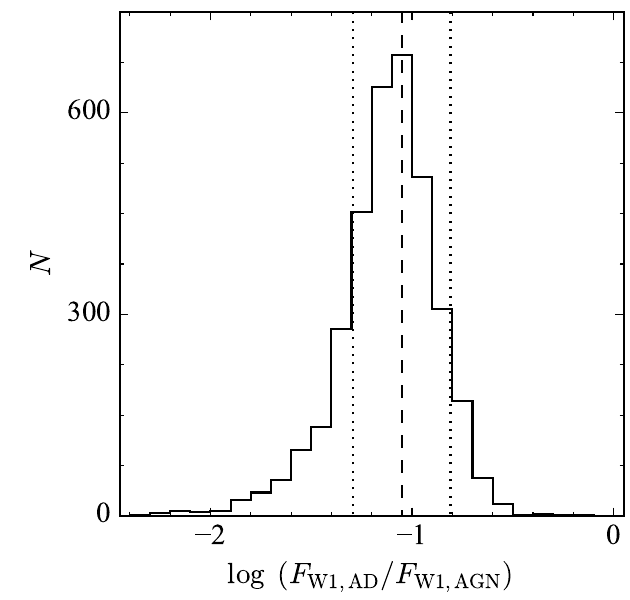}
\caption{
Distribution of the flux fraction of the continuum from the accretion disk in the W1 band, after subtracting host contribution, for type 1 AGNs. The expected flux from the accretion disk ($F_{\rm W1, AD}$) is calculated from the $g$-band flux. The dashed and dotted lines denote the mean value and standard deviation, respectively. 
}
\end{figure}

\subsection{Systematic Uncertainties}
As the sparse cadence, instrumental characteristics, and photometric errors can introduce systematic uncertainties into the SF, we perform extensive simulations to evaluate them. 

\subsubsection{Robust Structure Function Measurements}

There are two ways to estimate an ensemble SF: (1) the mean of $\sqrt{{\rm SF}^2}$ ($\overline{\rm SF}$) or (2) the square root of the mean of ${\rm SF}^2$ ($\sqrt{\overline{{\rm SF}^2}}$). Ideally, the two methods would yield similar results. However, the sparse cadence of the WISE dataset may introduce uncertainties in the ensemble SFs. We perform simulations with artificially generated light curves to test which method better reproduces the true SF. For simplicity, we assume that the light curves obey the damped random walk model with a break time scale of 650 days. The variability amplitude is assumed to be $10.6\%$ of the mean flux (SF$_\infty = 0.15$ mag), comparable to our sample. In total, $10^4$ light curves are generated randomly, considering the actual cadences and redshift distributions of our sample. We subsequently estimate the ensemble SF using the mock light curves. The photometric error is not included in this initial simulation. 

This experiment reveals that $\sqrt{\overline{{\rm SF}^2}}$ successfully reproduces the input SF, whereas the SF tends to be underestimated by $\overline{\rm SF}$ (Figure 2). More interestingly, for large $\Delta t$ ($\gtrsim 1000$ days), the discrepancy between the input SF and $\overline{\rm SF}$ becomes as severe as 20\%. This clearly indicates that $\sqrt{\overline{{\rm SF}^2}}$ needs to be used to trace the true SF of our sample. To account for the effect of photometric noise, we perform the same simulation by adding to the light curves a photometric noise of 0.017 mag, which is similar to the typical error in our sample. We again find that $\sqrt{\overline{{\rm SF}^2}}$ is more suitable than $\overline{\rm SF}$, but it deviates from the input value for $\Delta t \gtrsim 2500$ days. Unless otherwise noted, SF refers to $\sqrt{\overline{{\rm SF}^2}}$ throughout this paper.

\subsubsection{Photometric Errors}

As a particular field is visited every $\sim6$ months by NEOWISE, the same field is observed multiple times with $14\pm7$ single exposures in each visit. The photometry measurements are performed in a single exposure and combined into the representative magnitude in each epoch. Therefore, estimating the photometric noise ($\sigma_{\rm e}$) in this calculation is not straightforward. One way to compute the photometric noise is to utilize the SF of non-variable sources \cite[e.g.,][]{Kozlowski_2010}. For this purpose, we generate $10^4$ light curves with a constant flux associated with constant photometric noise ($\sigma_{\rm p}$), following the distributions of the cadences and redshifts of our sample. From the ensemble SF taken from these light curves, we find ${\rm SF} \approx \sqrt{2} \times \sigma_{\rm p}$, with SF independent of $\Delta t$. The $\sqrt{2}$ comes from the fact that $\sigma^2_p$ is subtracted twice, from $t$ and $t+\Delta t$ (see Equation 2).   

To mimic realistic observations, we perform the simulation using the NEOWISE multi-epoch data from non-variable sources. For this purpose, we randomly choose $10^4$ galaxies drawn from the set of inactive galaxies with neither AGN nor SF activity in the optical spectra contained in MPA-JHU catalogs by imposing the same redshift cut ($0.15 < z < 0.4$) as the AGN sample. 
As described in Section 2, we initially calculate the photometric noise ($\sigma_{\rm in}$) in each epoch using Equation (1). From this simulation, we find that the photometric uncertainty based on $\sigma_{\rm in}$ is significantly larger than the standard deviation of the magnitudes ($\sigma_{\rm rms}$) in the light curves of non-variable sources. Our experiment yields $\sigma_{\rm rms} \approx \sigma_{\rm in} / \sqrt{N_s}$, where $N_s$ denotes the number of single exposures in each epoch. From the SF measurements, we independently find that SF can be approximated by $\sqrt{2}\sigma_{\rm in} / \sqrt{N_s}$ (Figure 3), although it depends weakly on $\Delta t$, such that ${\rm SF} (\Delta t) \approx \sqrt{2} \sigma_{\rm in} / \sqrt{N_s}\times (0.999+0.055\Delta t/1000$. We adopt this relation when estimating the SF of our sample, noting that the dependence on $\Delta t$ barely affects the SF measurements. 

We note that the MIR continuum of inactive galaxies may be variable because of several possible origins, including supernova explosions, weak nuclear activity, or tidal disruption events, which can naturally introduce bias in the error estimation, even if their occurrence rates are relatively low \cite[e.g.,][]{jiang_2021, son_2022b, sun_2022}. If the inactive galaxies vary significantly in the MIR, our error estimation should be taken as an upper limit. To quantify this, we assume that the photometric noise can be described as $\sigma_{\rm in} / \sqrt{N_s}$ and rigorously estimate the ensemble SF of the inactive targets by subtracting this photometric noise using Equation (2). From this experiment, we found that the SF is $\sim 0.02$ mag regardless of $\Delta t$, which shows that the intrinsic variability of the inactive galaxies is almost negligible. Alternatively, if the inactive galaxies significantly vary in the MIR, the photometric errors are systematically smaller than our estimation (i.e., $\sigma < \sigma_{\rm in} / \sqrt{N_s}$). Even in this case, the photometric noise is substantially smaller than the SFs in our sample AGNs; therefore, its impact on the SF measurements is minimal.

\subsubsection{Uncertainties of the Ensemble Structure Function}
Finally, the uncertainty of the SF for an individual object is estimated from the standard deviation of each $\Delta t$ bin, while that of the ensemble SF is estimated by bootstrap resampling of the SFs for the individual galaxies. We create $10^3$ realizations of the SF for each object, and the final uncertainty is taken to be half of the difference between the 14th and 86th percentiles of the ensemble SF in each bin.

\begin{figure}[t!]
\centering
\includegraphics[width=0.45\textwidth]{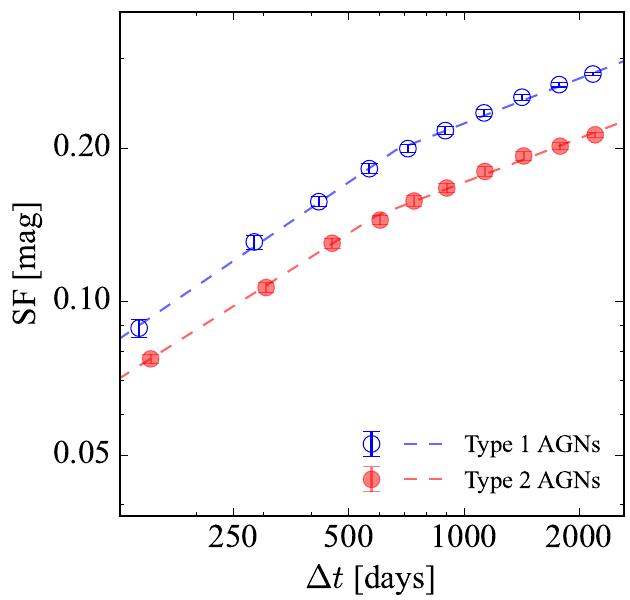}
\caption{
Ensemble SF of type 1 (open blue circles) and type 2 (filled red circles) AGNs. The fitting results with a broken power law are denoted by the blue and red dashed lines.    
}
\end{figure}

\subsection{Accretion Disk Contribution}

Although the W1 flux is generally dominated by the reprocessed emission from the hot and warm dust in the torus, the flux contribution from the AD may not be negligible for type 1 AGNs \cite[e.g.,][]{asmus_2011}. In addition, variability in the continuum emission from the AD can bias the SF. To properly remove the effect of the AD in the SF measurements, it would be ideal to subtract the contribution from the AD inferred from the corresponding optical photometric data in each epoch. Unfortunately, this is unavailable for our sample. Instead, we compute the expected W1 flux inferred from the SDSS $g$-band flux by assuming $F_\nu \propto \nu^{1/3}$, as predicted from theory and verified by polarization measurements of quasars \citep{vandenberk_2001, kishimoto_2008, li_2023}. Even after the host subtraction, the average flux contribution from the AD in the W1 band is $0.09\pm0.05$ (Figure 4). 

To estimate the expected SF at the W1 band solely due to the AD, we adopt the $g$-band SF of AGNs 
from \cite[][]{li_2023} that have a median \lbol\ $\approx 10^{45.3}$ erg s$^{-1}$ at $z<0.7$, similar to that of our sample. This SF is modeled with a variability amplitude SF$_\infty = 0.544$ mag, break time $\tau = 853.1$ days, and a power-law index $\beta\approx2\gamma = 1$ for small values of $\Delta t$. We also account for the dependence of SF on wavelength, $\rm SF \propto \lambda^{-0.30\pm0.05}$ \citep{ivezic_2004}, when converting the SF from $g$ to W1. We conservatively adopt an AD flux fraction of $0.17$, which is the 95th percentile value of the distribution of the AD fraction in the W1 band for our sample. The expected SF for $\Delta t = 1$ yr is $\sim0.032$, which indicates that the SF due to the AD is almost negligible. Therefore, we neglect this effect throughout the study.   

\begin{figure}[t!]
\centering
\includegraphics[width=0.45\textwidth]{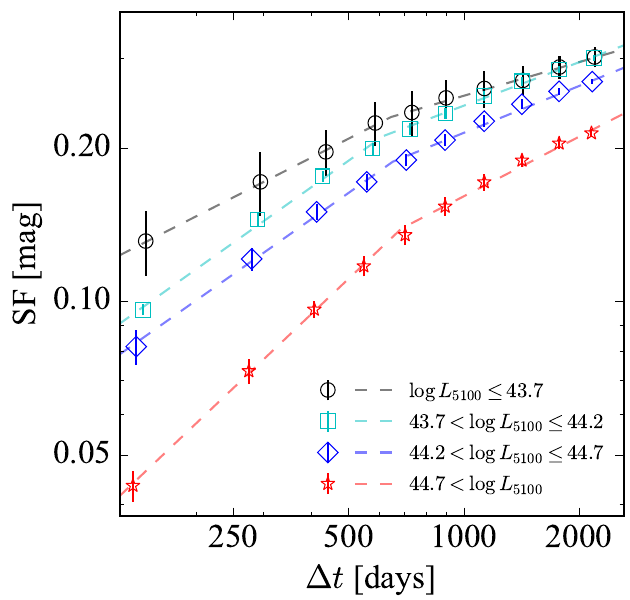}
\caption{
Ensemble SF for subsamples binned by the continuum luminosity at 5100~\AA, in units of erg s$^{-1}$.  The dashed lines denote the fits with a broken power law. 
}
\end{figure}

\begin{figure*}[t!]
\centering
\includegraphics[width=0.95\textwidth]{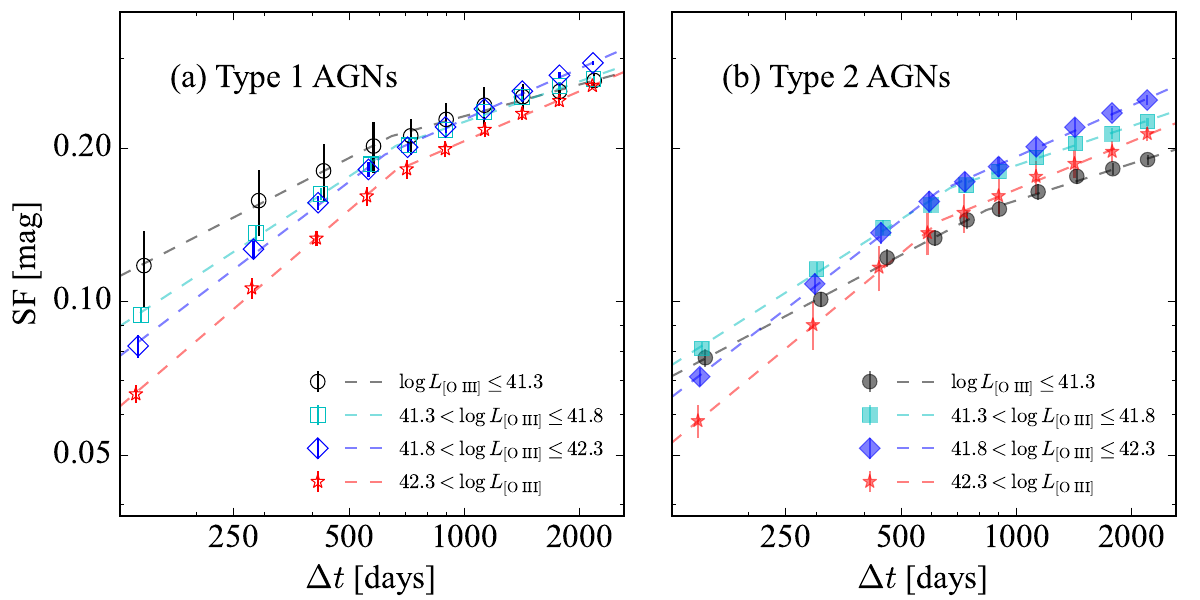}
\caption{
Ensemble SF of subsamples binned by \oiii\ luminosity, in units of erg s$^{-1}$. The open and filled symbols denote (a) type 1 and (b) type 2 AGNs, respectively. The dashed lines represent the fits with a broken power law. 
}
\end{figure*}

\section{Results}

We estimate the ensemble SFs for various subsamples according to AGN type, AGN property, and dust property. Previous studies have often modeled ensemble SFs of sparsely sampled light curves with a single power-law function (e.g., \citealt{kozlowski_2016b}). However, such a function is insufficient for describing the SFs in our sample. Instead, we fit the ensemble SF with a broken power-law function

\begin{equation}
SF (\Delta t) = 
\begin{cases}
A (\Delta t/365)^{\gamma_1} & \text{if } \Delta t \leq t_{\rm break} \\
A (t_{\rm break}/365)^{\gamma_1-\gamma_2}(\Delta t/365)^{\gamma_2} & \text{if } \Delta t > t_{\rm break}
\end{cases}
\end{equation}

\noindent
where $\Delta t$ is the time lag in days, $A$ is the variability amplitude at 1 yr, $t_{\rm break}$ is the break time scale, $\gamma_1$ is the power-law slope for $\Delta t \leq t_{\rm break}$, and $\gamma_2$ is the power-law slope at $\Delta t > t_{\rm break}$. For direct comparison between subsamples, we primarily consider the variability amplitude at 1 yr ($A$) and the slope for the lag at low $\Delta t$ ($\gamma_1$). We estimate the $1\sigma$ uncertainties of the fitted parameters through bootstrapping the ensemble SFs with their $1\sigma$ errors. On account of the sparse cadence of the WISE light curves, $t_{\rm break}$ is found to be sensitive to the initial conditions in the fitting of the ensemble SF, although its uncertainty appears to be small. Therefore, we decide not to overinterpret the trend of $t_{\rm break}$ in this study. The fitted parameters of the various subsamples are summarized in Table 1.

\begin{figure*}[t!]
\centering
\includegraphics[width=0.95\textwidth]{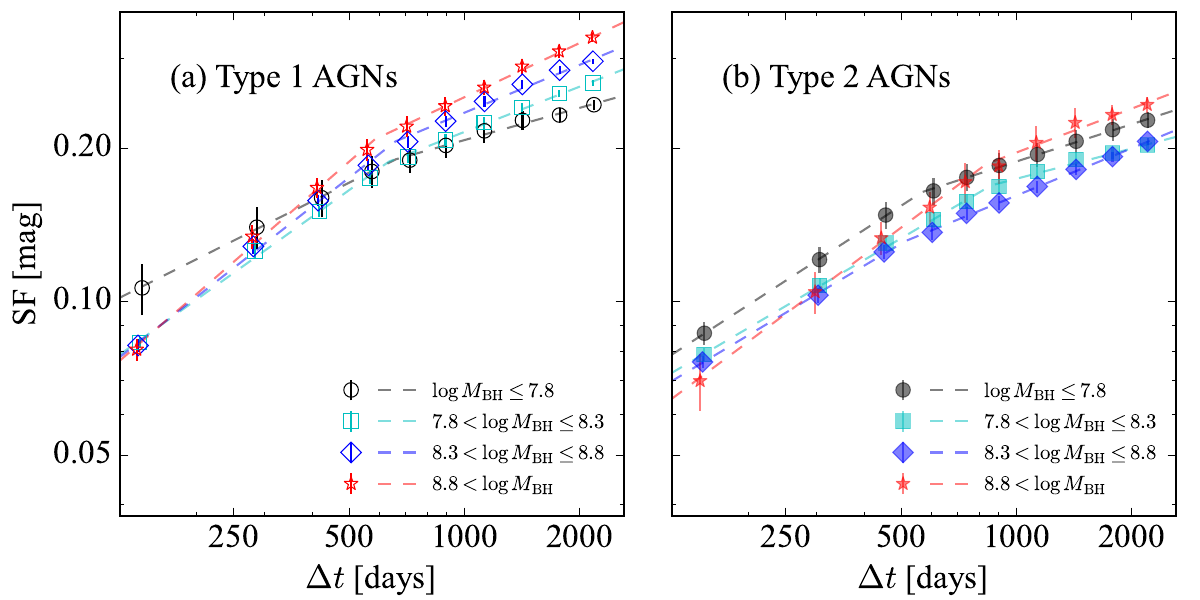}
\caption{
Ensemble SF of subsamples binned by BH mass, in units of $M_\odot$.  The open and filled symbols denote (a) type 1 and (b) type 2 AGNs, respectively. The dashed lines represent the fits with a broken power law. 
}
\end{figure*}

\begin{figure*}[t!]
\centering
\includegraphics[width=0.95\textwidth]{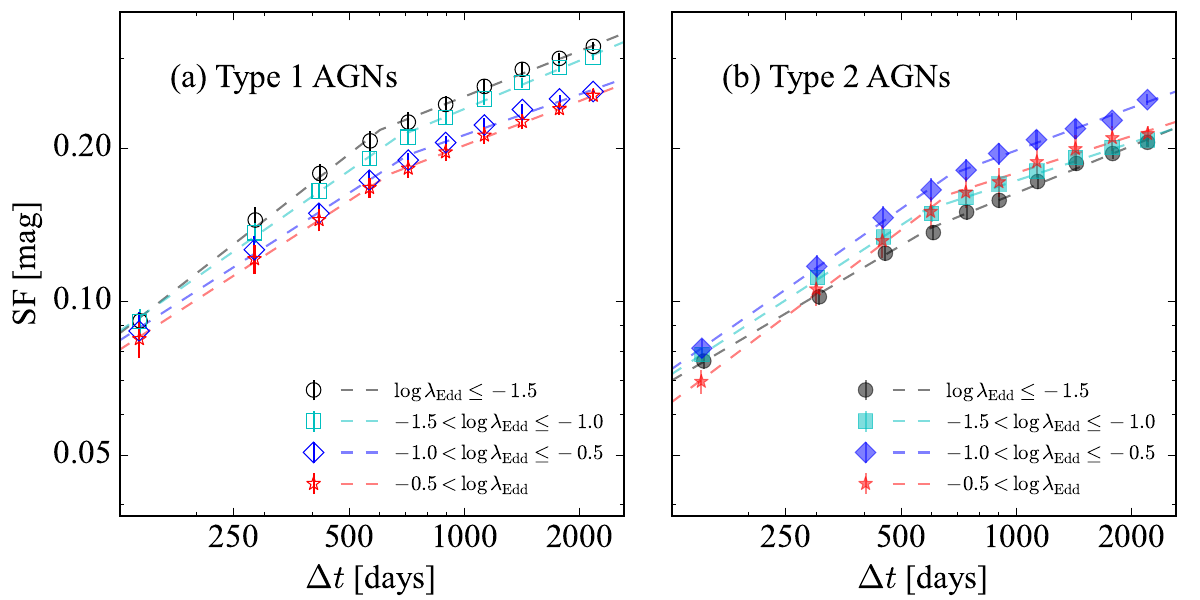}
\caption{
Ensemble SF of subsamples binned by Eddington ratio ($\lambda_{\rm Edd}$), using bolometric luminosities calculated from the \oiii\ luminosities.  The open and filled symbols denote (a) type 1 and (b) type 2 AGNs, respectively. The dashed lines represent the fits with a broken power law. 
}
\end{figure*}

\subsection{Type 1 vs. Type 2 AGNs}

According to the AGN unification model, type 1 and type 2 AGNs are determined by the viewing angle with respect to the opening angle of the dusty torus \citep{antonucci_1993, urry_1995}. We use the MIR SFs of type 1 and type 2 AGNs to test this model, which cannot be performed in the optical because the continuum emission from the AD is heavily obscured in type 2 AGNs. We find that the ensemble SFs of type 1 and type 2 AGNs are well fitted with a broken power-law model and are consistent in terms of their overall shape ($\gamma$), although type 2 AGNs exhibit a slightly shallower power law than their type 1 counterparts (Figure 5). On the other hand, the variability amplitude of type 1 AGNs is systematically larger than that of type 2 sources. These trends remain the same when the sample is divided into subsamples based on AGN properties. Interestingly, the similar power-law indices of type 1 and type 2 AGNs conflict with the results derived from optical SFs, for which the ensemble SFs of type 2 AGNs are significantly flatter than those of type 1 AGNs, possibly owing to obscuration \citep{decicco_2022}. Finally, our values of $\gamma_1$ ($0.47-0.51$) broadly agree with $\gamma \approx 0.45$ derived from the MIR (3.6 and 4.5 $\mu$m) light curves of high-$z$ quasars studied by \cite{kozlowski_2016a}.

\subsection{Dependence on AGN Properties}
The physical connection between the variability properties and AGN properties has been extensively studied with optical light curves.  We revisit this topic using MIR light curves. One of the most striking features in our sample is the dependence of the SF on the AGN luminosity. Using the monochromatic luminosity at 5100~\AA\ as a tracer of the bolometric luminosity for type 1 AGNs, Figure 6 clearly shows that low-luminosity AGNs tend to exhibit a larger variability amplitude ($A$) and a shallower slope ($\gamma_1$) than high-luminosity AGNs. The inverse correlation between $A$ and the AGN luminosity is consistent with that found for the optical SF \cite[e.g.,][]{kelly_2009, calpar_2017, sun_2018b, tang_2023}, which suggests that MIR variability may be directly related to the intrinsic variability of the optical light. The dependence of the slope on the luminosity is further discussed in Section 5.1.

To directly compare type 1 and type 2 AGNs, we adopt the \oiii\ luminosity as a proxy for the AGN luminosity. Although the overall trends remain the same, the dependence on the AGN luminosity as inferred from $L_{\rm [O\ III]}$ is slightly weaker compared to that derived using \l5100\ (Figure 7). This can be attributed to the less direct link between \oiii\ luminosity and the bolometric luminosity. Intriguingly, the ensemble SFs of type 2 AGNs of low-luminosity, defined here as $L_{\rm [O\ III]} \leq 10^{41.3}$ erg s$^{-1}$, significantly deviates from those of their type 1 counterparts, in the sense that the SF of low-luminosity type 2 AGNs tend to have lower amplitude and shallower slope.

We also examine how the ensemble SFs change with the BH mass and Eddington ratio (Figures 8 and 9). Although our results do not show any clear trend, $\gamma_1$ increases with increasing BH mass in type 1 AGNs. However, this trend is not clearly detected in type 2 AGNs, possibly owing to the relatively large uncertainties in the BH mass estimation. In addition, $A$ is weakly correlated with the Eddington ratio only in type 1 AGNs, presumably reflecting the dependence of the SF on AGN luminosity.

\begin{figure}[t!]
\centering
\includegraphics[width=0.45\textwidth]{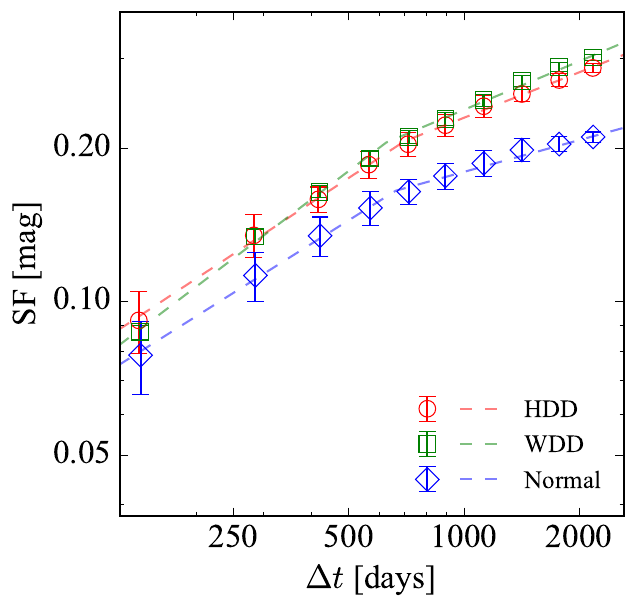}
\caption{
Ensemble SF for HDD (red circles), WDD (green squares), and normal (blue diamonds) AGNs. The fits with a broken power law are denoted by corresponding dashed lines.}
\end{figure}

\begin{figure*}[t!]
\centering
\includegraphics[width=0.95\textwidth]{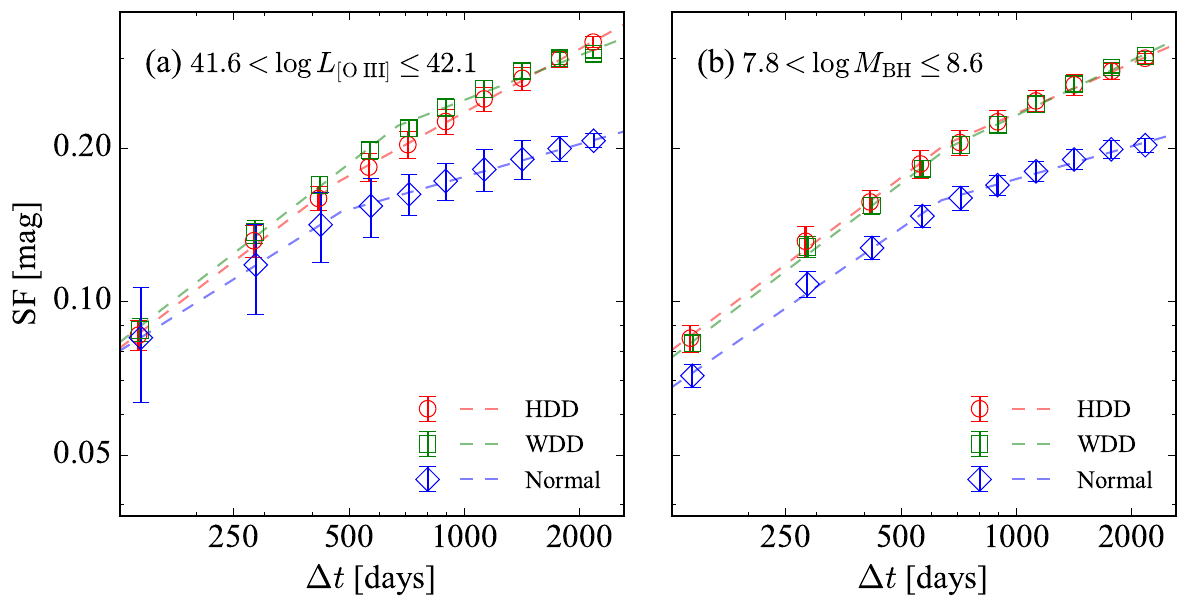}
\caption{
Ensemble SF for HDD (red circles), WDD (green squares), and normal (blue diamonds) AGNs at fixed (a) \oiii\ luminosity and (b) BH mass. The fits with a broken power law are denoted by corresponding dashed lines.}
\end{figure*}

\subsection{Dependence on Dust Properties}

Type 1 AGNs exhibit diverse MIR SEDs depending on the strength of the emission from their hot and warm dust components, which radiate mostly in the $1-3$ $\mu{\rm m}$ and $3-10$ $\mu{\rm m}$ regions, respectively. AGNs can be classified into three categories according to their SED shapes: normal, HDD, and WDD. We classify our sample of type 1 AGNs through SED fitting with empirically determined template spectra of the nearby quasars investigated by \cite{lyu_2017a}, who identified HDD and WDD AGNs as those whose MIR SEDs deviate from that of normal AGNs by more than $\ge0.3$ dex in the wavelength range $\sim 1-3$ $\mu{\rm m}$ and $\sim 3-10$ $\mu{\rm m}$, respectively. The fiducial reference SED for normal AGNs were taken from \citet{elvis_1994}. As described in detail in \citet{son_2023}, our fitting uses photometric data covering the $ugriz$ bands from SDSS, the $JHK_S$ bands from 2MASS, and the W1, W2, W3, and W4 bands from WISE. Using the Bayesian Information Criterion to determine the goodness of fit, we obtain best-fit models for 449 normal, 819 WDD, and 451 HDD AGNs.  

To test whether the dust properties are physically related to the MIR variability, we estimate the ensemble SFs for normal, WDD, and HDD AGNs. Figure 10 clearly demonstrates that the WDD and HDD AGNs exhibit larger variability amplitudes and steeper power-law slopes than normal AGNs. One caveat is that these findings could be a by-product of the dependence on AGN properties, as the dust properties are known to be closely connected with AGN parameters such as bolometric luminosity, Eddington ratio, and BH mass \cite[e.g.,][]{hao_2010, jiang_2010, jun_2013, lyu_2017a, son_2023}. To mitigate against this effect, we reestimate the ensemble SFs at a fixed, narrow range of \oiii\ luminosity and BH mass, but the dependence on dust properties still persists (Figure 11). We conclude that dust-deficient AGNs vary more drastically in the MIR than normal AGNs.

\section{Discussion}

Because the MIR SED of AGNs is dominated by thermal emission from the hot and warm dust in the torus, which is heated by the UV/optical photons from the AD, two separate components determine the MIR variability. On the one hand, the intrinsic variability of the flux from the AD should be reflected in the MIR variability. On the other hand, the MIR light curves are heavily smoothed compared to the optical light curves because the variability time scale associated with the torus is larger than that of the UV/optical continuum \cite[e.g.,][]{suganuma_2006, honig_2011, koshida_2014}. Consequently, the MIR SF can also be sensitive to the structure and physical properties of the torus. Here, we discuss the main drivers of the MIR variability based on the ensemble SFs of various subsamples.  

The most striking finding is the dependence of $A$ and $\gamma_1$ on the AGN luminosity. In particular, more luminous AGNs have systematically lower variability amplitudes (Figure 12), which agrees well with trends known in the optical (e.g., \citealt{vandenberk_2004, kelly_2009, tang_2023}; but see \citealt{sanchez_2018}). This may imply that there is a direct connection between the optical and MIR variability. By contrast, the dependence of the slope $\gamma_1$ for low $\Delta t$ on the AGN luminosity has not been identified in most previous studies involving optical light curves, suggesting that this trend may be attributed to the geometry of the torus.  

Interestingly, \citet{li_2023} demonstrated that the ensemble SFs of the MIR flux, in combination with those of the optical flux, can be used to estimate the inner radius of the dusty torus ($R_{\rm in}$). The MIR light curves can be represented by the convolution of the optical light curves with a transfer function of the torus, which depends on the torus geometry. In a toy model considering the torus geometry, \citet{li_2023} showed that the slope of the SF for low $\Delta t$ increases with increasing inner radius because the short-term variability in the optical light curves can be easily smoothed out with a large $R_{\rm in}$ (see Figure 4 in \citealp{li_2023}). In addition, the MIR variability can be substantially suppressed for large $R_{\rm in}$, as the torus transfer function is more widely spread over time at larger $R_{\rm in}$. Therefore, the inverse correlation between \oiii\ luminosity and $A$ ($\gamma_1$) may suggest that $R_{\rm in}$ is proportional to the AGN luminosity. Interestingly, this finding agrees well with the prediction from the size$-$luminosity relation, in which $R_{\rm in}$ is tightly correlated with the AGN luminosity \cite[e.g.,][]{koshida_2014, yang_2020, li_2023}. This finding is also in good agreement with the receding torus model \cite[e.g.,][]{barvainis_1987, lawrence_1991, simpson_2005}, in which $R{\rm in}$ is expected to be proportional to the AGN luminosity.

Meanwhile, some studies have argued that the slope $\gamma$ of the SF (${\rm SF} \propto \Delta t^\gamma$) is weakly correlated with the AGN luminosity even for the optical light curves (e.g., \citealt{kozlowski_2016b, smith_2018}). However, the variation of $\gamma$ in the optical studies is significantly smaller than that in the MIR, hinting that $\gamma$ derived from MIR variability is physically connected with the torus geometry. 

\begin{figure*}[t!]
\centering
\includegraphics[width=0.95\textwidth]{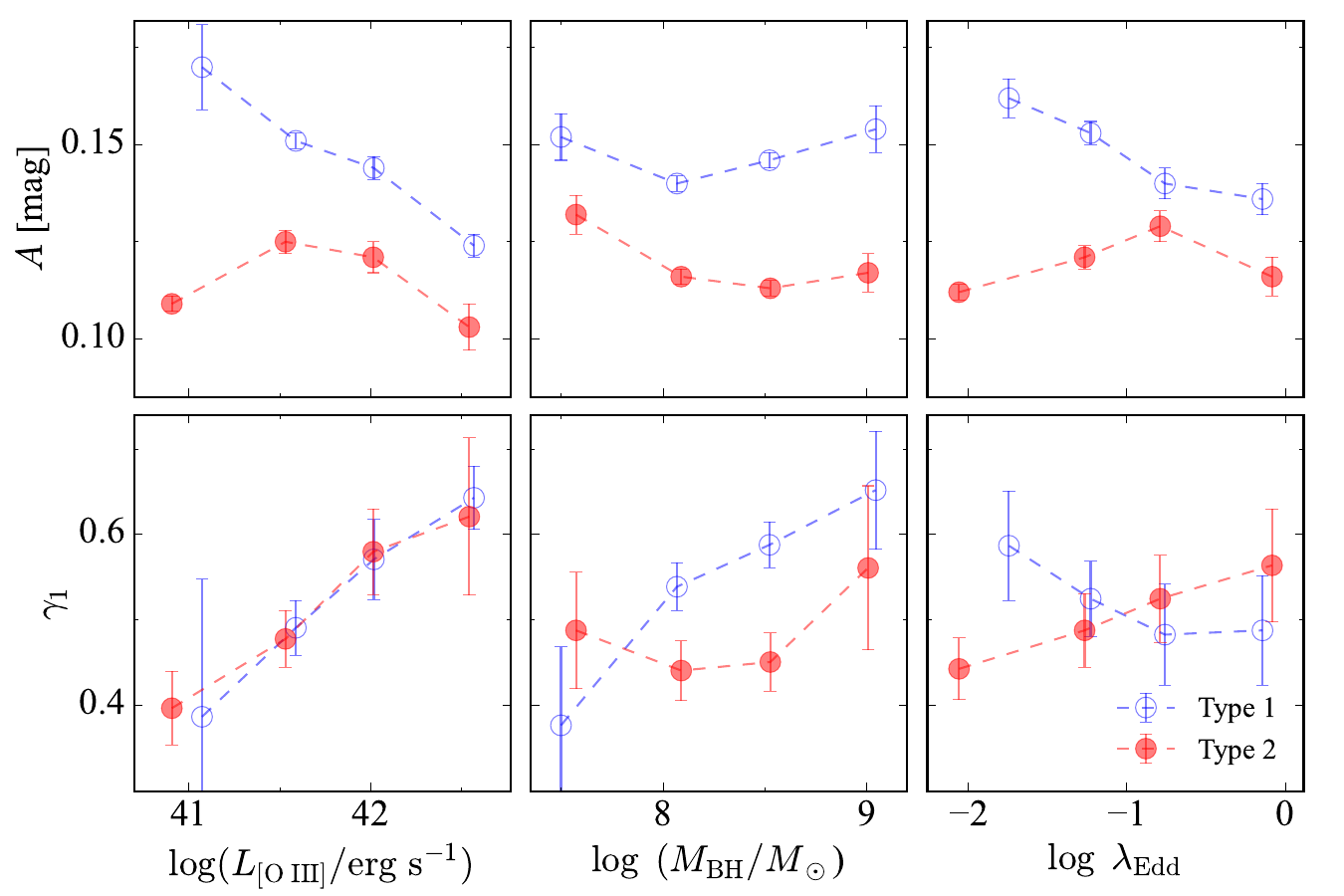}
\caption{
Dependence of $A$ (upper panels) and $\gamma_1$ (lower panels) on the \oiii\ luminosity (left), BH mass (middle), and Eddington ratio (right). The open blue and filled red circles denote type 1 and type 2 AGNs, respectively.
}
\end{figure*}

The distribution of $\gamma_1$ as a function of \oiii\ luminosity in type 1 and type 2 AGNs is almost identical (Figure 12). This suggests that both AGN types share a similar torus geometry, in line with expectation in the AGN unification paradigm. This result, however, seems to be at odds with the finding that type 2 AGNs have systematically lower variability amplitudes than type 1s. 
Extinction by the torus might be to blame, an explanation that resonates with previous observational and theoretical studies \citep{nenkova_2008, ramos_2011, hickox_2017}. Intriguingly, low-luminosity type 2 AGNs exhibit relatively low variability amplitudes, in apparent contradiction with type 1 AGNs and previous studies. The discrepancy is more evident over long-time scales.

While the physical origin of the variability behavior is unclear, we suspect that the W1 flux is contaminated by non-variable sources (e.g., warm and cold dust, or hot nanometer-sized grains in the host galaxies; \citealt{tran_2001, xie_2018}). Alternatively, the physical properties of the torus are distinctive from those of normal AGNs. In particular, the correlation between the variability properties and the Eddington ratio for type 1 and type 2 AGNs are clearly different (Figure 12). Some have argued that AGNs with very low Eddington ratios could not effectively form dense clouds around the AD \cite[e.g.,][]{elitzur_2009,elitzur_2014}. Such type 2 AGNs, so-called unobscured or intrinsic type 2 AGNs (\citealp{panessa_2002, tran_2011, ho_2012}), naturally lack a broad-line region and dusty torus. This may account for the lower detection rate of the broad-line region in LINERs than in normal AGNs \cite[e.g.][]{ho_1997, ho_2009}. According to this model, the low variability amplitude of low-luminosity type 2 AGNs can be attributed to the small torus covering factor, which poses a challenge to the traditional AGN unification model \citep{ichikawa_2015}. This may also reveal that the low $A$ in low-luminosity type 2 AGNs can be a by-product of the dependence on the Eddington ratio, as the bolometric luminosity is correlated with the Eddington ratio for type 2 AGNs (Figure 1). To test this scenario, we compare the \oiii\ luminosity and W1 luminosity ($L_{\rm W1}$) solely from the AGN, which is obtained from the SED fit described in Section 4.3 (Figure 13). For type 2 AGNs, we adopt the AGN templates with a large extinction ($A_v=20$) to account for the intrinsic extinction by the torus. We find that not only type 2 AGNs exhibit a lower fraction of $L_{\rm W1}$ to \oiii\ luminosity compared to type 1 AGNs, possibly because of the extinction, but the fraction also dramatically decreases toward low \oiii\ luminosity. This again implies that the covering factor of the torus in low-luminosity type 2 AGNs may be significantly lower than in type 1 AGNs of similar AGN luminosity.    

\begin{figure}[t!]
\centering
\includegraphics[width=0.45\textwidth]{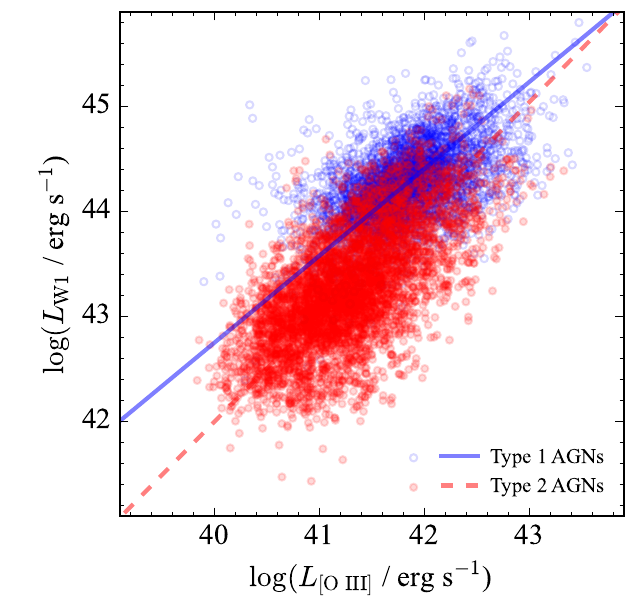}
\caption{
Correlation between \oiii\ luminosity and W1 luminosity, in units of erg s$^{-1}$. The symbols are the same as in Figure 1. The solid and dashed lines represent the linear fits for type 1 and type 2 AGNs, respectively.
}
\end{figure}

Dust-deficient AGNs systematically exhibit larger $A$, $\gamma_1$, and $\gamma_2$ compared to normal AGNs. As expected from the correlation between $R_{\rm in}$ and $\gamma_1$ inferred from the torus model of \citet{li_2023}, the steeper slopes for dust-deficient AGNs may indicate that the dust, which radiates the MIR continuum in dust-deficient AGNs, is likely distributed over larger radii than in normal AGNs \citep{li_2023}. This interpretation is consistent with the fact that dust-deficient AGNs lack hot and warm dust, which is more centrally located within the torus. In addition, a larger variability amplitude in dust-deficient AGNs than normal AGNs regardless of AGN properties may imply that the hot and warm dust is physically unstable in dust-deficient AGNs. Perhaps it can be easily destroyed and reformed, depending on the strength of the high-energy photons from the AD, which would naturally lead to a high variability amplitude in the MIR. Alternatively, the systematic differences between normal and dust-deficient AGNs in terms of their MIR SF may arise from the fact that the two types intrinsically exhibit different variability characteristics in the UV/optical continuum from the AD. This can be further addressed in future studies with optical light curves.

\section{Summary}

We assemble a $10$-year long multi-epoch MIR dataset for type 1 and type 2 AGNs selected from SDSS that are also observed by NEOWISE in the W1 band.  With these data, we examine the MIR variability by rigorously computing the ensemble SF of subsamples according to various AGN properties. We also perform an extensive simulation to consider the sparse cadence of the WISE dataset, investigate the systematic uncertainty involved, and evaluate the uncertainties in the SF measurements. This analysis yields the following results.

\begin{itemize}
\item The overall shapes of the ensemble SFs for both type 1 and type 2 AGNs are nearly identical, except that the MIR variability amplitude of type 1 AGNs is systematically larger than that of type 2 sources. This may be attributed to the large extinction in type 2 AGNs.   

\item The variability amplitude is inversely correlated with the AGN luminosity, likely a reflection of the intrinsic variability of the light from the AD. The same trend holds for the slope of the SF for low $\Delta t$ ($\gamma_1$), which we interpret as the positive correlation between the inner radius of the torus and the AGN luminosity.

\item Dust-deficient AGNs tend to have larger variability amplitudes and slopes in the ensemble SF than do normal AGNs. While the physical origin of the larger variability amplitudes and slopes remains unclear, one factor may be that the covering factor of the centrally located hot and warm dust sensitively responds to the variation of UV photons from the AD.

\item Low-luminosity type 2 AGNs tend to have lower variability amplitudes than their type 1 counterpart or high-luminosity type 2 AGNs. We suggest that low-luminosity type 2 AGNs have a characteristically low torus covering factor. Alternatively, the W1 flux is significantly contaminated by the cold dust in the host galaxy.
\end{itemize}

The above results lead us to conclude that MIR variability can be a powerful tool for probing the torus properties of AGNs. For example, time-series near-infrared spectroscopic data obtained through the upcoming space mission Spectro-Photometer for the History of the Universe, Epoch of Reionization and Ices Explorer (SPHEREx) will play a crucial role in this effort \cite[e.g.,][]{dore_2018, kim_2021b}.

\begin{acknowledgments}
We thank the anonymous referee for constructive comments that greatly helped to improve the manuscript.
This work was supported by the National Science Foundation of China (11721303, 11991052, 12011540375, 12233001), the China Manned Space Project (CMS-CSST-2021-A04, CMS-CSST-2021-A06), and the National Research Foundation of Korea (NRF), through a  grant funded by the Korean government (MSIT) (No. 2022R1A4A3031306 and 2023R1A2C1006261).
\end{acknowledgments}

\medskip

\begin{rotatetable*}
\begin{deluxetable*}{cccccccccccc}
\tablecolumns{12}
\tablenum{1}
\tabletypesize{\scriptsize}
\tablewidth{0pc}
\tablecaption{Structure Function Parameters\label{tab:table1}}
\tablehead{
\colhead{\h} &
\multicolumn{5}{c}{\h Type 1 AGNs} &
\colhead{\h} &
\multicolumn{5}{c}{\h Type 2 AGNs} \\
\cline{2-6}
\cline{8-12} 
\colhead{\h Subsample} &
\colhead{\h $N$} &
\colhead{\h $A$} &
\colhead{\h $\Delta t_{\rm break}$} &
\colhead{\h $\gamma_1$} &
\colhead{\h $\gamma_2$} &
\colhead{\h} &
\colhead{\h $N$} &
\colhead{\h $A$} &
\colhead{\h $\Delta t_{\rm break}$} &
\colhead{\h $\gamma_1$} &
\colhead{\h $\gamma_2$} \\
\colhead{\h} &
\colhead{\h} &
\colhead{\h (mag)} &
\colhead{\h (days)} &
\colhead{\h} &
\colhead{\h} &
\colhead{\h} &
\colhead{\h} &
\colhead{\h (mag)} &
\colhead{\h (days)} &
\colhead{\h} &
\colhead{\h} \\
\colhead{\h (1)} &
\colhead{\h (2)} &
\colhead{\h (3)} &
\colhead{\h (4)} &
\colhead{\h (5)} &
\colhead{\h (6)} &
\colhead{\h} &
\colhead{\h (7)} &
\colhead{\h (8)} &
\colhead{\h (9)} &
\colhead{\h (10)} &
\colhead{\h (11)}
}
\startdata
All &\h 3506 &\h $0.146\pm0.002$ &\h $660\pm102$ &\h $0.513\pm0.031$ &\h $0.293\pm0.018$ &\h &\h 3074 &\h $0.117\pm0.002$ &\h $585\pm130$ &\h $0.476\pm0.026$ &\h $0.287\pm0.023$ \\
 &\h      &\h       &\h     &\h           &\h               &\h               &\h &\h      &\h               &\h           &\h               \\
$\log L_{5100} \leq 43.7$&\h 209 &\h $0.186\pm0.012$ &\h $641\pm293$ &\h $0.383\pm0.146$ &\h $0.219\pm0.072$ &\h &\h \nd  &\h \nd           &\h \nd       &\h \nd           &\h \nd   \\
$43.7 < \log L_{5100} \leq 44.2$&\h 1340 &\h $0.160\pm0.003$ &\h $612\pm101$ &\h $0.535\pm0.033$ &\h $0.283\pm0.022$ &\h &\h \nd  &\h \nd           &\h \nd       &\h \nd           &\h \nd   \\
$44.2 < \log L_{5100} \leq 44.7$&\h 1630 &\h $0.139\pm0.003$ &\h $649\pm147$ &\h $0.530\pm0.060$ &\h $0.305\pm0.026$ &\h &\h \nd  &\h \nd           &\h \nd       &\h \nd           &\h \nd   \\
$44.7 < \log L_{5100}$&\h 326 &\h $0.089\pm0.002$ &\h $684\pm156$ &\h $0.713\pm0.055$ &\h $0.388\pm0.051$ &\h &\h \nd  &\h \nd           &\h \nd       &\h \nd           &\h \nd   \\
 &\h      &\h       &\h     &\h           &\h               &\h               &\h &\h      &\h               &\h           &\h               \\
$\log L_\mathrm{[O\ III]} \leq 41.3$&\h 360 &\h $0.170\pm0.011$ &\h $642\pm282$ &\h $0.387\pm0.161$ &\h $0.206\pm0.065$ &\h &\h 1465 &\h $0.109\pm0.002$ &\h $829\pm293$ &\h $0.397\pm0.043$ &\h $0.240\pm0.048$ \\
$41.3 < \log L_\mathrm{[O\ III]} \leq 41.8$&\h 1275 &\h $0.151\pm0.002$ &\h $660\pm125$ &\h $0.491\pm0.032$ &\h $0.263\pm0.028$ &\h &\h 1055 &\h $0.125\pm0.003$ &\h $669\pm134$ &\h $0.478\pm0.033$ &\h $0.256\pm0.031$ \\
$41.8 < \log L_\mathrm{[O\ III]} \leq 42.3$&\h 1334 &\h $0.144\pm0.003$ &\h $617\pm99$ &\h $0.571\pm0.047$ &\h $0.332\pm0.022$ &\h &\h 433 &\h $0.121\pm0.004$ &\h $620\pm134$ &\h $0.580\pm0.050$ &\h $0.334\pm0.035$ \\
$42.3 < \log L_\mathrm{[O\ III]}$&\h 519 &\h $0.124\pm0.003$ &\h $655\pm108$ &\h $0.643\pm0.037$ &\h $0.325\pm0.032$ &\h &\h 121 &\h $0.103\pm0.006$ &\h $631\pm204$ &\h $0.621\pm0.092$ &\h $0.311\pm0.063$ \\
 &\h      &\h       &\h     &\h           &\h               &\h               &\h &\h      &\h               &\h           &\h               \\
$\log M_\mathrm{BH} \leq 7.8$&\h 803 &\h $0.152\pm0.006$ &\h $658\pm294$ &\h $0.377\pm0.092$ &\h $0.210\pm0.053$ &\h &\h 275 &\h $0.132\pm0.005$ &\h $566\pm143$ &\h $0.488\pm0.068$ &\h $0.243\pm0.047$ \\
$7.8 < \log M_\mathrm{BH} \leq 8.3$&\h 1235 &\h $0.140\pm0.002$ &\h $632\pm76$ &\h $0.539\pm0.028$ &\h $0.295\pm0.023$ &\h &\h 1044 &\h $0.116\pm0.002$ &\h $857\pm235$ &\h $0.441\pm0.035$ &\h $0.197\pm0.057$ \\
$8.3 < \log M_\mathrm{BH} \leq 8.8$&\h 1090 &\h $0.146\pm0.002$ &\h $658\pm102$ &\h $0.588\pm0.027$ &\h $0.312\pm0.025$ &\h &\h 1332 &\h $0.113\pm0.002$ &\h $507\pm122$ &\h $0.451\pm0.034$ &\h $0.308\pm0.021$ \\
$8.8 < \log M_\mathrm{BH}$&\h 361 &\h $0.154\pm0.006$ &\h $582\pm134$ &\h $0.652\pm0.069$ &\h $0.353\pm0.035$ &\h &\h 423 &\h $0.117\pm0.005$ &\h $831\pm298$ &\h $0.561\pm0.096$ &\h $0.286\pm0.112$ \\
 &\h      &\h       &\h     &\h           &\h               &\h               &\h &\h      &\h               &\h           &\h               \\
$\log \lambda_\mathrm{Edd} \leq -1.5$&\h 469 &\h $0.162\pm0.005$ &\h $600\pm131$ &\h $0.587\pm0.064$ &\h $0.297\pm0.033$ &\h &\h 1733 &\h $0.112\pm0.002$ &\h $591\pm200$ &\h $0.443\pm0.036$ &\h $0.307\pm0.033$ \\
$-1.5 < \log \lambda_\mathrm{Edd} \leq -1.0$&\h 1031 &\h $0.153\pm0.003$ &\h $665\pm145$ &\h $0.525\pm0.044$ &\h $0.315\pm0.030$ &\h &\h 744 &\h $0.121\pm0.003$ &\h $564\pm135$ &\h $0.488\pm0.043$ &\h $0.247\pm0.035$ \\
$-1.0 < \log \lambda_\mathrm{Edd} \leq -0.5$&\h 1182 &\h $0.140\pm0.004$ &\h $712\pm226$ &\h $0.483\pm0.059$ &\h $0.265\pm0.040$ &\h &\h 391 &\h $0.129\pm0.004$ &\h $666\pm169$ &\h $0.525\pm0.051$ &\h $0.277\pm0.043$ \\
$-0.5 < \log \lambda_\mathrm{Edd}$&\h 795 &\h $0.136\pm0.004$ &\h $639\pm160$ &\h $0.488\pm0.064$ &\h $0.289\pm0.032$ &\h &\h 206 &\h $0.116\pm0.005$ &\h $661\pm178$ &\h $0.564\pm0.066$ &\h $0.240\pm0.062$ \\
 &\h      &\h       &\h     &\h           &\h               &\h               &\h &\h      &\h               &\h           &\h               \\
HDD&\h 451 &\h $0.150\pm0.006$ &\h $694\pm199$ &\h $0.496\pm0.091$ &\h $0.296\pm0.047$ &\h &\h \nd  &\h \nd           &\h \nd       &\h \nd           &\h \nd  \\
WDD&\h 819 &\h $0.151\pm0.003$ &\h $624\pm97$ &\h $0.570\pm0.033$ &\h $0.316\pm0.025$ &\h &\h \nd  &\h \nd           &\h \nd       &\h \nd           &\h \nd  \\
Normal&\h 449 &\h $0.125\pm0.007$ &\h $661\pm254$ &\h $0.473\pm0.124$ &\h $0.207\pm0.058$ &\h &\h \nd  &\h \nd           &\h \nd       &\h \nd           &\h \nd
\enddata
\tablecomments{
SFs for each subsample.
Col. (1): Subsample.
\ll5100: logarithmic 5100~\AA\ continuum luminosity in units of erg s$^{-1}$;
\lloiii: logarithmic \oiii\ luminosity in units of erg s$^{-1}$;
\llmbh: logarithmic BH mass in the units of $M_{\odot}$;
HDD: hot dust-deficient AGNs; WDD: warm dust-deficient AGNs; Normal: normal AGNs.
Col. (2): Subsample size.
Col. (3): SF amplitude at $\Delta t = 1$ yr.
Col. (4): Break in the broken power law.
Col. (5): Power-law index at $\Delta t < \Delta t_{\rm break}$.
Col. (6): Power-law index at $\Delta t > \Delta t_{\rm break}$.
}
\end{deluxetable*}
\end{rotatetable*}

\facilities{IRSA}

\smallskip

\software{Astropy \citep{astropy_2013,astropy_2018,astropy_2022}, Scipy \citep{scipy_2020}}

\bibliography{torus}

\end{document}